\documentclass[12pt]{iopart}

\usepackage{iopams}  
\usepackage{bm}
\usepackage{amstext}
\usepackage{setstack}
\usepackage{graphicx}
\begin{document}

\title{Magnetic moment of an electron near a surface with dispersion}

\author{Robert Bennett and Claudia Eberlein}

\address{Department of Physics \& Astronomy, University of Sussex, Falmer, Brighton BN1 9QH, UK}

\begin{abstract}
Boundary-dependent radiative corrections that modify the magnetic moment of an electron near a dielectric or conducting surface are investigated. 
Normal-mode quantization of the electromagnetic field and perturbation theory applied to the Dirac equation for a charged particle in a weak magnetic field yield a general formula for the magnetic moment correction in terms of any choice of electromagnetic mode functions. For two particular models, a non-dispersive dielectric and an undamped plasma, it is shown that, by using contour integration techniques over a complex wave vector, this can be simplified to a formula featuring just integrals over TE and TM reflection coefficients of the surface. Analysing the magnetic moment correction for several models of surfaces, we obtain markedly different results from the previously considered simplistic `perfect reflector' model, which is due to the inclusion of physically important features of the surface like evanescent field modes and dispersion in the material. Remarkably, for a general dispersive dielectric surface, the magnetic moment correction of an electron nearby has a peak whose position and height can be tuned by choice of material parameters. 
\end{abstract}

\maketitle

The spin magnetic moment of an electron differs from the value predicted by the Dirac equation because the electron is coupled to the quantized electromagnetic field. The electromagnetic field is in turn modified by the presence of a material boundary; thus the magnetic moment for an electron near a surface differs from the value found in free space \cite{Fischbach, Svozil, Bordag, BoulwareBrown, KreuzerSvozil, Kreuzer, BartonFawcett}. This shift is, in principle, measurable spectroscopically, however previous investigations have shown that this effect is so minute as to be judged immeasurable with the then attainable experimental precision. All these previous works \cite{Fischbach, Svozil, Bordag, BoulwareBrown, KreuzerSvozil, Kreuzer, BartonFawcett} model the surface as a perfect reflector, which forces the parallel electric and the perpendicular magnetic fields on the surface to vanish and can be represented by a medium with infinite refractive index. While simple to work with, this model has some obvious physical deficiencies in that it ignores electromagnetic field modes that are evanescent outside the medium and in that the medium does not become transparent at infinite frequencies as any real material would. For some quantum electrodynamic surface effects like the Casmir-Polder force on an atom in front of a surface \cite{CP},
the idealized model of a perfect reflector works just fine and its results are reproduced as limiting cases e.g. of a non-dispersive dielectric medium but with infinite refractive index \cite{Wu} or a dispersive plasma surface but with infinite plasma frequency $\omega_p$ \cite{BabikerBarton}. This Letter is going to show that the situation is radically different for the spin magnetic moment of an electron near a surface where different models for the electromagnetic response of the surface give quite different results, not necessarily obtainable as limiting cases of each other. That this can happen was already found for the inertial mass shift of a free electron near a surface \cite{EberleinRobaschik}, the reason for which was found to be connected to the fact that a free particle admits excitations of arbitrarily low frequency, thus interacting with photons of arbitrarily long wavelengths which are dealt with very differently in dielectric and perfect-reflector models of a reflecting surface.

In order to determine the quantum-field theoretical corrections to the electron's magnetic moment, one usually calculates the vertex diagram to the respective order of interest. To 1-loop order $e^2\equiv\alpha$ and in free space, this is a straightforward calculation and graduate textbook material (c.f. e.g. \cite{Peskin}), but when the electromagnetic field is reflected by a surface even 1-loop calculations of quantum electrodynamics get rather complicated \cite{PRDEberleinRobaschik}, largely because of the loss of translation invariance and the requirement to localize the particle in relation to the surface. Here we are interested only in the {\em correction} to the magnetic moment due to the presence of the material surface nearby, whence we take a different, more appropriate approach that dispenses with quantizing the electron field and just works with the quantized photon field and first-quantized electrons. This works because the Feynman graphs that contribute to 1-loop order are the irreducible vertex correction and reducible graphs of the bare vertex with 1-loop self-energy and vacuum-polarization insertions (cf. e.g. Fig. 12.15 of \cite{LeBellac}), and of those the vacuum polarization is the only contribution that would necessitate the second-quantization of the electron. However, the vacuum polarization does not, to 1-loop order, contain any internal photon lines and hence to this order does not receive any boundary-dependent corrections. Thus, we can work out the boundary-dependent part of the shift by quantizing just the electromagnetic field and use perturbation theory in the Dirac equation to determine the energy shift $-\boldsymbol{\mu}\cdot{\bf B}_0$, due to the presence of the material surface, of the electron's spin in a weak external magnetic field ${\bf B}_0$. Alternatively, one could of course take a non-relativistic expansion of the Dirac equation first, by means of a Foldy-Wouthuysen transformation or otherwise, and then use perturbation theory to work out the same energy shift, as has been done for working out the magnetic moment correction of the electron near a perfect reflector \cite{BartonFawcett}. This approach requires care since several successive orders in the non-relativistic expansion turn out to contribute to the shift \cite{BartonFawcett}. By comparison, the present paper's calculation using perturbation theory in the Dirac euqation is much more straightforward and physically transparent as well as less error prone.

We start by considering a quantized electromagnetic field interacting with an electron that sits near a half-space of constant refractive index $n$. Initially we shall assume that the weak classical magnetic field ${\bf B}_0\equiv B_0 \hat{z}$ interacting with the magnetic moment is normal to the surface spanning the $x y$ plane, so that in the resulting energy shift the coefficients of terms linear in $\sigma_z B_0$ yield the spin magnetic moment. 
We write the quantized electromagnetic field in terms of mode functions $\mathbf{f}_{\mathbf{k}\lambda}$ and standard photon creation and annihilation operators as
\begin{equation}
\boldsymbol{A} = \sum_{\text{all modes}} \hspace{-8mm}\int\hspace{3mm} (\mathbf{f}_{\mathbf{k}\lambda} a_{\mathbf{k} \lambda}+\mathbf{f}_{\mathbf{k}\lambda}^* a_{\mathbf{k} \lambda}^\dagger) ,
\label{Afield}
\end{equation}
where $\lambda$ denotes a particular polarization, and any normalization constants have been absorbed into the mode functions. We then apply perturbation theory to the Dirac equation for an electron coupled to the electromagnetic field via minimal coupling by using the Dirac electron's states in a constant magnetic field \cite{JohnsonLippmann} as unperturbed states and the coupling term $-e\gamma_0\boldsymbol{\gamma}\cdot\mathbf{A}$ with the quantized photon field as perturbation. Second-order perturbation theory gives relative magnetic moment corrections $\Delta\mu/\mu$ of order $e^2$. The non-relativistic expansion we make in the process eventually turns out to lead to a series in powers of the ratio of the electron's Compton wavelength and its distance $z$ from the surface, $1/mz$ \footnote{We work in natural units $c = 1 = \hbar$, $\epsilon_0 = 1$.}. The leading order boundary-dependent correction to the magnetic moment is of order $\Delta\mu/\mu\approx e^2/(mz)^2$, which in fact relates to another main reason beside technical complications for using an approach based on the quantized photon field interacting with a first-quantized Dirac electron over a fully second-quantized quantum electrodynamic approach: as shown by Kreuzer \cite{Kreuzer} for a perfectly reflecting surface, the quantum electrodynamic approach gives leading-order contributions to the $g$ factor as well as to the mass shift that are of order $e^2/(mz)$ but which cancel in the boundary-dependent corrections to the magnetic moment. The same happens if dielectric models of the surface are used \cite{unpubl}, and thus the derivation of the boundary-dependent magnetic-moment corrections requires the notoriously difficult calculation of next-to-leading-order terms in $1/mz$. By contrast, perturbation theory applied to the Dirac equation, as we use here, or to the Schr\"odinger equation, as used in Ref.~\cite{BartonFawcett}, circumvents any such problems because one evaluates the magnetic moment shift directly and the leading terms are of order $e^2/(mz)^2$ from the outset.

The application of perturbation theory is straightforward, if somewhat lengthy, with the one exception that the dipole approximation commonly used in this approach well-known from quantum optics applications is not universally applicable. In some terms one needs to go beyond the dipole approximation by not just taking the value of the quantum field $\mathbf{A}$ at the position of the electron but by Taylor-expanding it around this position and treating the resulting powers of the displacement operator $\mathbf{r}-\mathbf{r}_0$ as operator acting on the Landau states of the electron in the external magnetic field, which generates additional factors of $B_0^{-1/2}$. 
The end result turns out to be independent of the Landau level. We find
\begin{eqnarray}
\fl\mu_\perp =-\frac{e^3}{4m^3}\sum \Bigg[{ |f_z|^2}+\frac{|{(\nabla \times \mathbf{f})}_x|^2}{\omega^2} + \frac{|{(\nabla \times \mathbf{f})}_y|^2}{\omega^2} \nonumber\\
\hspace*{20mm}
+\frac{1}{\omega^2}\left(f_x\frac{\partial^2f_y^* }{\partial x \partial y}+ f_y\frac{\partial^2f_x^*}{\partial x\partial y} -f_y\frac{\partial^2f_y^*}{\partial x^2}  -f_x\frac{\partial^2f_x^*}{\partial y^2}     + \text{C.C.} \right)\Bigg],   \label{ShiftModeFunctions}
\end{eqnarray}
which, we emphasize, is a completely general result for any quantized photon field of the form (\ref{Afield}). For a non-dispersive dielectric half-space the mode functions can be found in Ref.~\cite{CarnigliaMandel}. Substituting these into the above expression, and renormalizing the shift by subtracting the value it would take in free space, we find that if we write the photon momentum as $\mathbf{k} = \mathbf{k}_\parallel + k_z \hat{z}$ we can deform the contour in the complex $k_z$ plane to give the results in terms of double integrals over the reflection coefficients at the surface. By cycling the Cartesian coordinates in the mode functions we can also consider surfaces parallel to the external field without much additional effort and thus give the results for the shifts along both directions:
\begin{eqnarray}
\fl \Delta \mu_\perp = \frac{e^3}{16 \pi^2  m^3} \left\{\int_0^\infty d\xi \xi \int^\infty_1 d\eta \left[ \left( 3\eta^2-2\right)R_{TE}+\left(\eta^2-2 \right)\left(R_{TM}-\frac{\epsilon(0)-1}{\epsilon(0)+1}\right)\right] e^{2\xi \eta z} \right. \nonumber\\ \left. \hspace*{7mm}
- \frac{\epsilon(0)-1}{\epsilon(0)+1}\; \frac{3}{4  z^2 } \right\} \label{FullResultPerp} \\
\fl \Delta \mu_\parallel = \frac{e^3}{16 \pi^2  m^3} \left\{\frac{1}{2}\int_0^\infty d\xi \xi \int^\infty_1 d\eta \left[ \left( \eta^2-3\right)R_{TE}+\left(5\eta^2-3 \right)\left(R_{TM}-\frac{\epsilon(0)-1}{\epsilon(0)+1}\right)\right] e^{2\xi \eta z} \right. \nonumber\\ \left. \hspace*{7mm}
- \frac{\epsilon(0)-1}{\epsilon(0)+1}\; \frac{1}{ z^2 } \right\} \label{FullResultPara}
\end{eqnarray}
\begin{equation}
\fl \text{with}  \qquad R_{TE} = \frac{\eta -\sqrt{(\epsilon(\xi)-1)+\eta^2}}{\eta+\sqrt{(\epsilon(\xi)-1)+\eta^2}} 
\quad, \quad  R_{TM} = \frac{\eta \epsilon(\xi)-\sqrt{(\epsilon(\xi)-1)+\eta^2}}{\eta\epsilon(\xi)+\sqrt{(\epsilon(\xi)-1)+\eta^2}}\ .
\label{Rcoeffs}
\end{equation}
Here $\xi=-i\omega$ originates from the photon frequency and $\eta=k_z/\omega$ is the cosine of the angle of incidence. For later convenience, we have written the dielectric function $\epsilon(\xi)$ as a quantity that could depend on frequency, but for the non-dispersive case we simply have $\epsilon(\xi) = n^2 = \epsilon(0)$. The terms outside the integrals come from the subtraction and separate consideration of the point at $k_z = -ik_\parallel$, where TM modes contribute an electrostatic term, which is mathematically expedient.
For the non-dispersive dielectric these integrals can be evaluated analytically and yield:
\begin{eqnarray}
\fl\Delta \mu_\perp =-\frac{e^3}{32\pi ^2m^3 z^2}\frac{1}{\left(n^4-1\right)^{3/2} }\Bigg[\sqrt{n^4-1} (5-2 n+n^2-2 n^3-3 n^4+n^5)  \nonumber\\ 
-n^4 \sqrt{n^2-1} \left(1+2 n^2\right) \text{arctanh}\left(\frac{(n-1) \sqrt{1+n^2}}{1+(n-1) n}\right)\nonumber\\ +2 \left(n^2-1\right) \left(1+n^2\right)^{5/2} \text{ln}\left(n+\sqrt{n^2-1}\right)\Bigg] \\
\fl\Delta \mu_\parallel =-\frac{e^3}{192\pi ^2m^3 z^2}\frac{1}{\left(n^4-1\right)^{3/2} }\Bigg[\sqrt{n^4-1}(26-9 n+8 n^2-23 n^3-3 n^4+n^5) \nonumber\\
+3 n^4 \sqrt{n^2-1} \left(2-3 n^2\right)  \text{arctanh}\left(\frac{(n-1) \sqrt{1+n^2}}{1+(n-1) n}\right)\nonumber\\ +9 \left(n^2-1\right) \left(1+n^2\right)^{5/2} \text{ln}\left(n+\sqrt{n^2-1}\right)\Bigg]
\end{eqnarray}
For large refractive indices the reflection coefficients become those of a perfect reflector, whence one might expect that in the limit $n\rightarrow\infty$ the above results for $\Delta\mu$ would reproduce those previously derived for an electron near a perfectly reflecting surface \cite{BartonFawcett}. However, in that limit we find
\begin{equation}
\Delta\mu_\perp = -\frac{e^2}{4\pi}\frac{e}{2 m} \left(\frac{n}{4 \pi m^2 z^2}-\frac{1}{4 \pi m^2 z^2}+\mathcal{O}(1/n)\right) \label{Largennondisp}
\end{equation}
\begin{equation}
\Delta\mu_\parallel =-\frac{e^2}{4\pi}\frac{e}{2 m} \left(\frac{n}{24 \pi m^2 z^2}+\frac{1}{4 \pi m^2 z^2}+\mathcal{O}(1/n)\right)\;. \label{LargennondispRotated}
\end{equation}
The linear rise with $n$ of the leading terms appears to be unphysical, as this would suggest that the magnetic moment could be increased arbitrarily by increasing the refractive index $n$ of the surface. However, as we shall explain below, this misconception arises from the unphysical assumption of a dispersionless medium. A curious observation to note is that the next-to-leading terms independent of $n$  in Eqs. (\ref{Largennondisp}) and (\ref{LargennondispRotated}), if taken on their own, do in fact reproduce the results of the perfect-reflector case (cf. Eq. (7.12) of \cite{BartonFawcett}). A consistency check of taking the limit $n\rightarrow\infty$ in the reflection coefficients (\ref{Rcoeffs}) first and evaluating the integrals (\ref{FullResultPerp}) and (\ref{FullResultPara}) afterwards reproduces the results of Ref.~\cite{BartonFawcett} and reveals the mathematical origin of the discrepancy: the limits of infinite refractive index, $n\to\infty$, and $\eta \to \infty$, which corresponds to the static limit $k_z \to -ik_\parallel$, in the TE reflection coefficient do not commute. As we shall explain later on this is an indication of the decisive role played by the low-frequency character of the dielectric susceptibility of the material that makes up the surface.

To shed light on the problem, we consider the interaction of the an electron with a dispersive surface. We start with the simplest model of a dispersive medium, a plasma surface \cite{ElsonRitchie} with the dielectric function
\begin{equation}
\epsilon(\omega) = 1- \frac{\omega_p^2}{\omega^2}
\end{equation}
where $\omega_p$ is the  plasma frequency. 
The normal modes $f_{{\bf k}\lambda}$ for this system can be derived following Refs.~\cite{ElsonRitchie,BabikerBarton}, though for the present purpose it is expedient to separate left- and right-incident modes and write the modes in the same form as for the non-dispersive medium \cite{CarnigliaMandel}. On substitution of these into Eq.~(\ref{ShiftModeFunctions}) we find that manipulating the contour in the complex $k_z$ plane (noting in particular that the contribution from a pole in the TM reflection coefficient cancels with the contribution from the surface plasmons, as seen in \cite{BabikerPlasma}) gives precisely Eqs.~(\ref{FullResultPerp}) and (\ref{FullResultPara}), just with a different dielectric function entering into the reflection coefficients. However, for this particular dielectric function there is a technical problem with the evaluation of the integrals for the TE modes. For this case we in fact deform the contour back down to the real $k_z$ axis and evaluate the integral directly. This is relatively straightforward because the TE reflection coefficient for a plasma surface takes a particularly simple form. The magnetic moment shifts we obtain then agree in the limit $\omega_p \to \infty$ with the perfect-mirror results given by \cite{BartonFawcett}, but not with the limit $n\to \infty$ of the non-dispersive dielectric model, as explained above. This is not altogether surprising since neither the plasma nor the perfect-reflector models permit modes that are travelling inside the medium and evanescent on the vacuum side, which, however, feature in the dielectric model. We also find the at first sight worrying result that as we send $\omega_p\to 0$ the shift in the plasma model diverges, while one would expect it to be zero. Mathematically this arises because the $\omega_p\to 0$ and $k_z \to -i k_\parallel$ limits of the TM reflection coefficient do not commute, indicating that the static polarizability of the medium plays a major role in the shift. The nature of this discrepancy is explained further in Ref.~\cite{massshift}.

\begin{figure}
\includegraphics{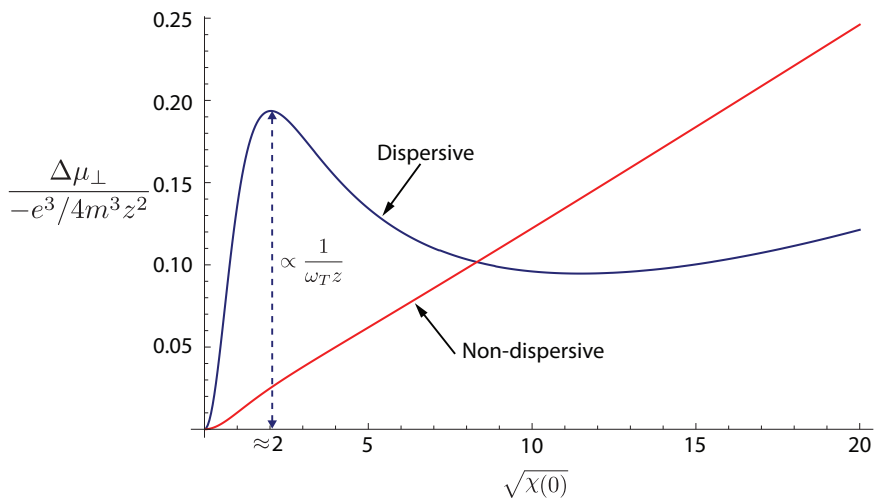}
\caption{\label{fig:graph} $\Delta\mu_\perp$ for a dispersive surface with $|\omega_T z| = 0.02$ and for a non-dispersive surface}
\end{figure}

To resolve all these issues we turn to a dispersive dielectric as model for the surface. Along the lines of \cite{BartonPlasma} we introduce a restoring force into the equation of motion governing the dielectric response of the material, which leads to the dielectric function
\begin{equation}
\epsilon(\omega) = 1- \frac{\omega_p^2}{\omega^2-\omega_T^2}\;.
\label{eps-diel}
\end{equation}
This model now includes total internal reflection inside the medium and evanescent modes in vacuum, so we expect results with features in common with the non-dispersive dielectric.
A derivation of the magnetic moment shift from first principles is now much more complicated, as the equation for the field modes $f_{{\bf k}\lambda}$ cannot be written as an Hermitean eigenvalue problem. Instead one would need to include both dispersion and absorption into the model by constructing a Huttner-Barnett-type field theory for the dielectric \cite{RJZarxiv}. A short-cut is provided by the Lifshitz theory, where just as for atomic energy level shifts near surfaces \cite{WylieSipe}, for an end result that depends only on the surface's reflection coefficients, one necessarily gets the same expressions (\ref{FullResultPerp}) and (\ref{FullResultPara}), as before, for the magnetic moment shifts. Substituting the dielectric function (\ref{eps-diel}) into Eqs.~(\ref{FullResultPerp}--\ref{Rcoeffs}), we find that for a dispersive dielectric the magnetic moment shift exhibits a peak at certain values of the material parameters. In order to show how the shift could be maximized by a specific choice of these parameters, we plot
in Fig.~\ref{fig:graph} the magnetic moment shift near a dispersive dielectric as a function of the static dielectric susceptibility $\chi(0) = \epsilon(0)-1$, which equals 
$\omega_p^2/\omega_T^2$ for a dispersive dielectric and $n^2-1$ for the non-dispersive model.
If we were to continue the plot to large values of $\chi(0)$, the graphs for the two models converge into one linearly rising graph. The peak is well-defined at values of $|\omega_T z| \lesssim 0.07$ and $|\omega_T z|\lesssim 0.25$ for ${\bf B}_0$ perpendicular and parallel to the surface, respectively, but disappears for larger values of $|\omega_T z|$. The ratio of the height of the peak to the non-dispersive result at the same $\chi(0)$ is given by
\begin{equation}
\frac{\Delta\mu_{\perp {\rm disp}}}{\Delta\mu_{\perp{\rm nondisp}}} \approx \frac{30.3 \,\text{eVnm}}{|\omega_T z|}\;, \quad 
\frac{\Delta\mu_{\parallel {\rm disp}}}{\Delta\mu_{\parallel {\rm nondisp}}} \approx \frac{81.6\,\text{eVnm}}{|\omega_T z|}\;.
\end{equation}
As $|\omega_T z|$ decreases, we find that the peak moves closer to $\sqrt{\chi(0)}\equiv\omega_p/\omega_T  \approx 2$. Typical values of $\omega_T$ for metals are on the order of a few eV (see, for example, \cite{SiliconDielectricFunction}), meaning that an extremely small $z$ would be required to get a significant enhancement. However, materials can be tuned to have $\omega_T$  significantly smaller than this - for example an InSb semiconductor grating can have $\omega_T$ (and $\omega_p$) in the range of a few meV \cite{SemiconductorPlasma}. For $z$ of a few tens of nanometres this corresponds to an enhancement factor on the order of $10^3$. 

The large $\chi(0)$ behaviour of the result sheds light on the apparent problem of the non-dispersive result in the limit $n\to\infty$. For large $\chi(0)$, we observe that the shift for the dispersive dielectric model becomes linear in $\sqrt{\chi(0)}$ and agrees with the non-dispersive results; so for large $\chi(0)$ the two models are equivalent. However, an arbitrarily large $\chi(0)$ would model a dielectric with an arbitrarily large static polarizability which is physically not possible. The comparison with the plasma and perfect reflector models shows that the physics is different for those: in the plasma model the static limit is infinite right from the start, whence results do agree with the perfect reflector in the limit $\omega_p \to \infty$. 

Therefore, we conclude that the magnetic moment shift in a given set-up crucially depends on the choice of a physically accurate model for the low-frequency behaviour of the material making up the surface. In particular, it matters whether the material is a conductor or an insulator, i.e.~whether the charge carriers inside the material are freely mobile or subject to a restoring force by the ion cores. Mathematically, the discrepancies between the various results reported above come from non-commutation of certain limits of the reflection coefficient, namely a non-commutation between the static limit ($k_z \to -ik_\parallel$) and whichever limit we have to take in order to compare models. We have already discussed the $n\to\infty$ limit of infinite refractive index in the non-dispersive model. 
Similarly, the limit $\omega_T\to 0$ (leading to $\chi(0) \to \infty$) in the dispersive dielectric model does not reproduce the results of the plasma model because the limits $\omega_T\to 0$ and $k_z \to -ik_\parallel$ of the dispersive TE reflection coefficient do not commute.

The shift for the non-dispersive dielectric model has a distance dependence of $1/z^2$ for all $n$. However, in the plasma model we find a distance dependence of $1/z^3$ for $\omega_p z\ll 1$, namely:
\begin{equation}
\Delta \mu_\perp (|\omega_p z| \ll1) \approx \frac{e^3}{64 \pi \sqrt{2}\omega_p m^3 z^3} \label{smallomegapzperp}
\end{equation} 
\begin{equation}
\Delta \mu_\parallel (|\omega_p z| \ll1) \approx \frac{5e^3}{128 \pi \sqrt{2}\omega_p m^3 z^3} \label{smallomegapzparallel}
\end{equation} 
This is because at small distances the dominant effect is the electrostatic interaction of the magnetic moment with the surface plasmon and the $1/z^3$ terms come entirely from the surface plasmon part of the mode functions, similarly to what is discussed in Ref.~\cite{BabikerPlasma}. For the same reasons, the dispersive dielectric model, which features surface polaritons, also yields a $1/z^3$ dependence of the shift in the limit $\omega_p z\ll 1$.

When discussing the measurability of the effect, one has to carefully consider both the set-up and the experimental methods used. Our analysis shows that the relative 
magnetic moment shift $\Delta\mu/\mu$ can be of order of $10^{-9}$ for an electron that is 10nm away 
from a dispersive dielectric surface, which compares very favourably with the current experimental accuracy of $10^{-12}$ for $g/2$ in free space \cite{Odom}, though for a distance of 100nm from the surface the prediction for $\Delta\mu/\mu$ shrinks to just $10^{-11}$ which is only barely above current experimental accuracy in free space. However, quite apart from whether one could realistically perform such a precision experiment very close to a surface, one needs to carefully consider how the measurement is performed. Since one usually cannot measure the strength of the applied magnetic field $B_0$ to any high accuracy, experiments tend to not measure the magnetic moment directly but rather its ratio to either a known magnetic moment or to the cyclotron frequency of the particle, as in Ref.~\cite{Odom}. For such an experiment carried out close to a surface, one would need to take into account the shift in cyclotron frequency of the particle caused by the surface, which arises due to the self-energy of the particle being dependent on the distance from the surface \cite{BartonFawcett,EberleinRobaschik,massshift}. The self-energy shift of a Dirac particle is linear in the ratio of the Compton wavelength to the distance from the surface, i.e.~it goes with $1/(mz)$, so that it is many orders of magnitude bigger than the shift in the magnetic moment, which goes with $1/(mz)^2$. In other words, such an experiment would effectively measure the self-energy shift of the particle, which would be interesting in its own right, but not the magnetic moment shift.

In summary, we have shown that the magnetic moment shift near a dispersive dielectric surface is notably different from the predictions made for a perfectly reflecting surface. For the case of the magnetic field ${\bf B}_0$ normal to surface even the sign is different than predicted by a perfect-reflector model.
Additionally, we have shown that the inclusion of dispersion can significantly modify the magnitude of the effect and its distance dependence, and that these modifications can be tuned by choice of material parameters. A physically realistic choice of model for the low-frequency behaviour of the dielectric function is essential for obtaining physically correct results. Our formulae (\ref{FullResultPerp})--(\ref{Rcoeffs}) can be applied to get the magnetic moment shifts near a surface provided the frequency-dependence of the reflection coefficients is known.

It is a pleasure to thank Robert Zietal for discussions. Financial support from the UK Engineering \& Physical Sciences Research Council is gratefully acknowledged.

\section*{References}
\bibliography{NJPMagneticMoment}
\bibliographystyle{unsrt}

\end{document}